\documentclass{eas}
\usepackage{graphicx}
\TitreGlobal{The Role of Disk--Halo Interaction in Galaxy Evolution: 
Outflow vs Infall?}
\runningtitle{Interaction of HVCs with the Outskirts of Galactic Disks}
\begin{document}

\title{Interaction of HVCs with the Outskirts of Galactic Disks: Turbulence} 
\author{A. Santill\'an}\address{Direcci\'on General de Servicios de C\'omputo
Acad\'emico, UNAM, 04510, Mexico City, Mexico}
\author{F. J. S\'anchez--Salcedo}\address{Instituto de Astronom\'\i a, UNAM, 
04510, Mexico City, Mexico}
\author{J. Kim}\address{Korea Astronomy Observatory, 61--1, Hwaam--Dong, 
Yusong--Ku, Taejon 305--348, Korea}
\author{J. Franco}\sameaddress{2}
\author{L. Hern\'andez--Cervantes}\sameaddress{2}
\begin{abstract}
There exist many physical processes that may contribute to the driving of 
turbulence in galactic disks. 
Some of them could drive turbulence 
even in the absence of star formation. For example, hydrodynamic (HD) or 
magnetohydrodynamic (MHD) instabilities, frequent mergers of small satellite 
clumps, ram pressure, or infalling gas clouds. In this work we present 
numerical simulations to study the interaction of compact high 
velocity clouds (CHVC) with the outskirts of magnetized gaseous disks. With 
our numerical simulations we show that the rain of small HVCs onto the disk 
is a potential source of random 
motions in the outer parts of H{\sc i} disks.
\end{abstract}
\maketitle
\section{Introduction}
There is solid evidence that, in most 
spiral galaxies, the linewiths of H\,{\sc i} emission line, $\sigma$, 
vary radially from $\sim 12$ to $15$ km s$^{-1}$ in 
the central parts to a very constant value between $6$ to $8$ km s$^{-1}$ in 
the outer parts (\cite{Dib2006}). 
Our main goal is to assess how much of the velocity dispersion observed 
in extended H\,{\sc i} galactic disks is due to the impact of HVCs and 
intermediate velocity clouds (IVCs). 
Strong evidence for the presence of continuing gaseous infall
to the Galactic disk and external galaxies has been compiled 
in \cite{bec03}.
Not all the accreting material should have an extragalactic origin.
\cite{booth07} showed that
the Galactic fountain can efficiently cycle matter from the center
of the Galaxy to its outskirts at a rate of $\sim 0.5$ M$_{\odot}$ yr$^{-1}$.

\section{Numerical Model and Results}

The simulations were performed with the MHD code ZEUS--3D (Stone \& Norman 
1992a, 1992b) and the MHD--TVD code (\cite{kim99}). 
Our local frame of reference is rotating with the disk. 
at a distance $R_0 \simeq 20$ kpc from the galactic center. 
We have run two-dimensional (2D) and three-dimensional (3D) simulations.
The coordinates ($x$,$y$) correspond to the 
horizontal (planar) direction and $z$ to the perpendicular 
direction. 
The boundary conditions are {\it periodic} in the $x$--axis and
$y$--axis, and {\it outflow} 
in the $z$--axis. The evolution was computed in the quasi--isothermal regime
($\gamma$=1.01), ignoring gas self--gravity and differential rotation.
In the 2D simulations,
the magnetic field has only one component along the $y$--axis and is initially 
stratified in the $z$--direction, $B_{\rm y}$($z$). 
The ISM is initially  in magnetohydrostatic equilibrium. 
The resolution of the domain for our 2D simulations was of 
$1024\times 1024$ zones, and $128\times 128 
\times 256$ zones for the 3D case. 

All infalling clouds have the same physical characteristics, 
initial radius $R_{cl} = 50$~pc and internal density 
$n_{cl} = 0.1$~cm$^{-3}$. The accretion mass rate due to the
accreting clouds is $0.5$ M$_{\odot}$ yr$^{-1}$.
The CHVCs are injected at $z_{cl}=\pm 2$~kpc,
i.e., the injection occurs in both caps of the disk, with
a vertical velocity $v_{cl} = 100$~km~s$^{-1}$.
The case where the injection occurs only along one side 
was discussed in Santill\'an et al.~(2007).

\begin{figure}
\includegraphics[height=5.3in,width=5.3in]{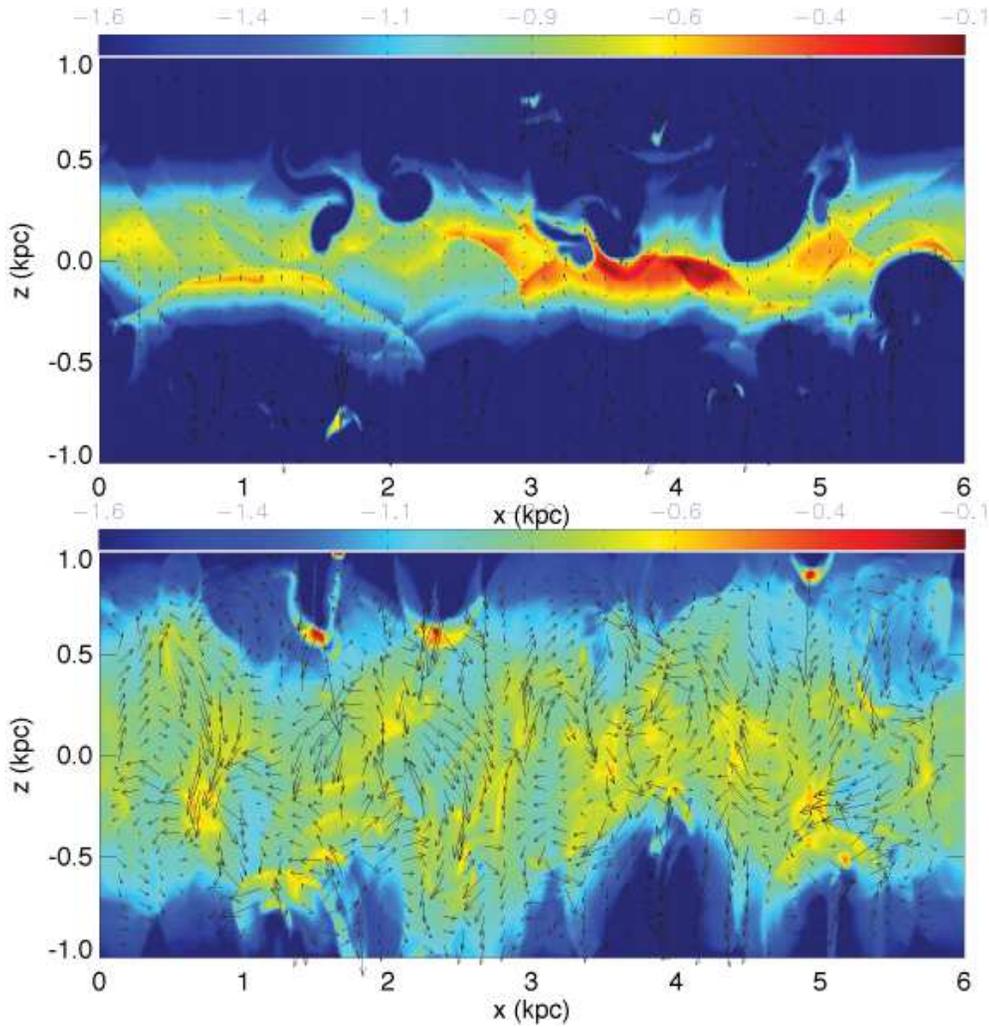}
  \caption{Evolution of the gas in the outer parts of the galactic disks. 
The sequence shows the density (\textit{color logarithmic scale}) and velocity 
field (\textit{arrows}) at two selected times (0.1 and 2~Gyr).}\label{fig:hvc}
\end{figure}


\begin{figure}
\includegraphics[width=6cm]{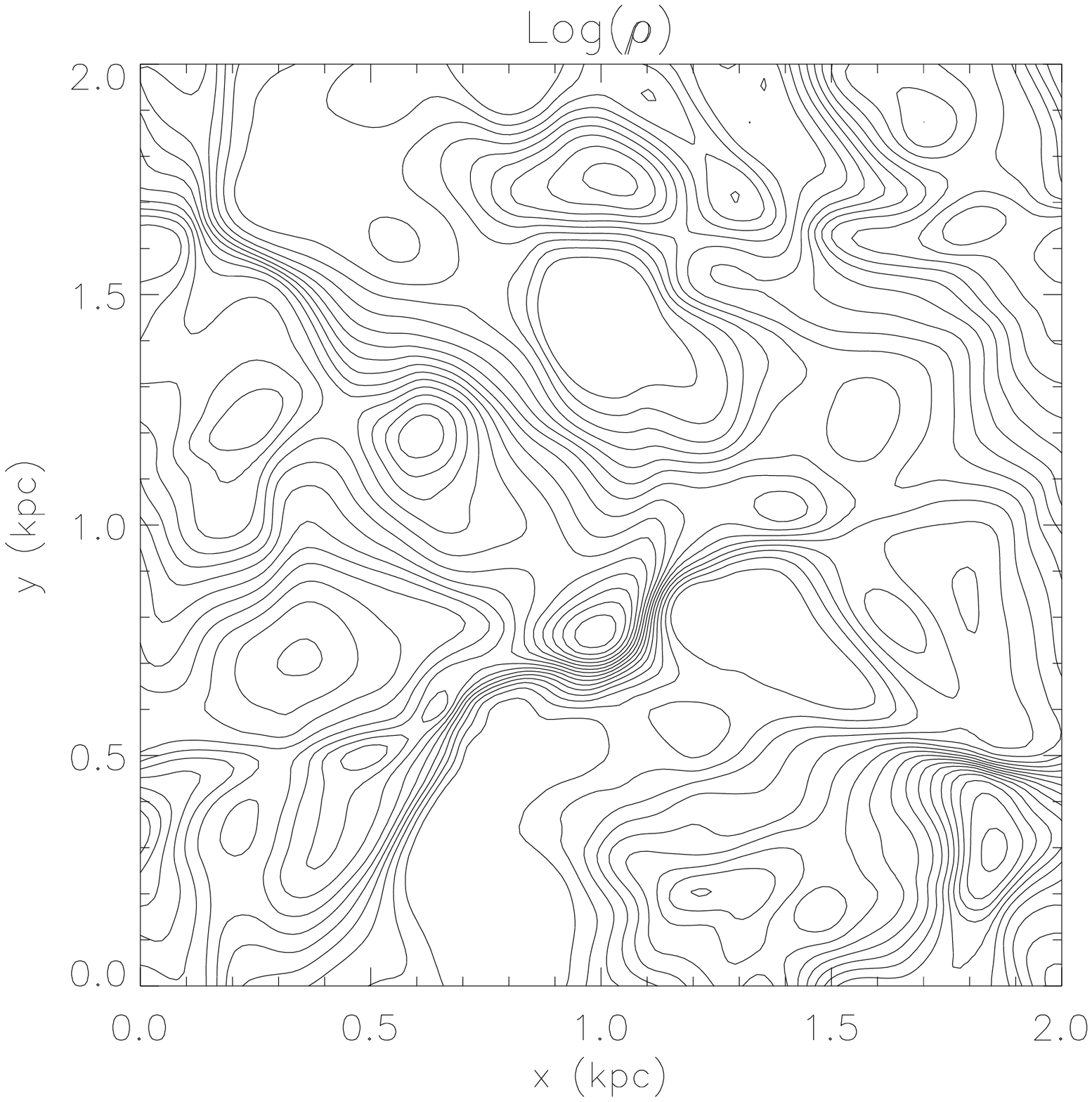}
\qquad
\includegraphics[width=6cm]{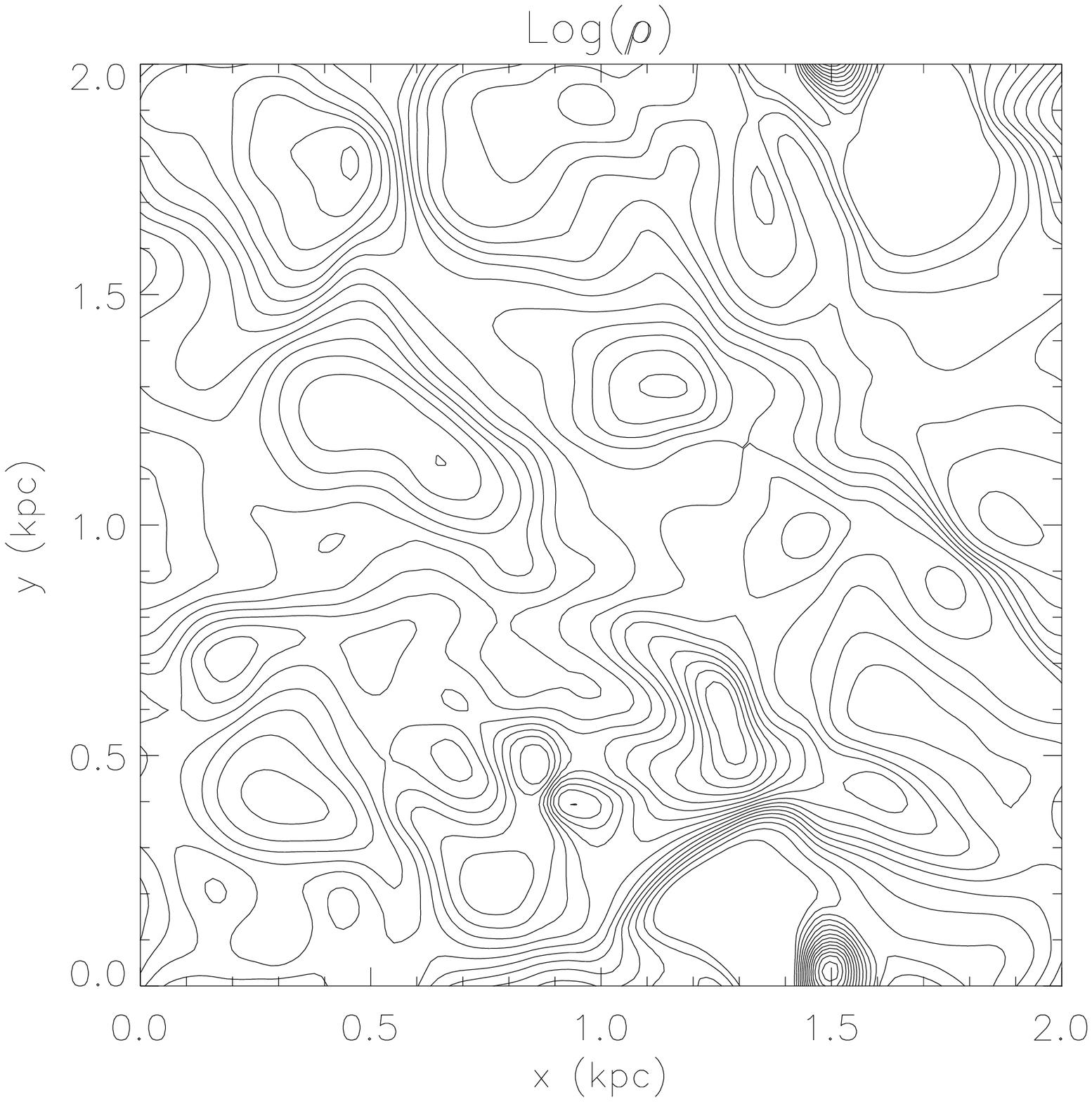}
\caption{Density contours in the $z=0$ plane at two selected times (0.375 
and 0.400 Gyr).}\label{fig:planexy}
\end{figure}


The 2D evolution of the ISM in the outskirts of a Galactic disk is shown 
in Figure~\ref{fig:hvc}.
The disk suffers the continuous bombardment 
of CHVCs during 2~Gyr. The impact of each cloud produces a 
strong galactic shock directed downward and a reverse shock that penetrates 
into the cloud (\cite{santi99}). 
When two shocks that moves in opposite directions collide, most of
the kinetic motion, which ultimately contributes to the gas velocity dispersion,
is dissipated. 
The steady-state vertical velocity dispersion is close to transonic as long as
the injection velocity of the clouds is $>50$ km s$^{-1}$.
In the vertical
direction, accelerated gas due to incident clouds are decelerated by
the external gravitational potential and falls back to the disk.
Again, this falling material may collide with ascending gas, leading
to energy dissipation and momentum cancellation. 

Although the details differ, the basic gasdynamic phenomena described above 
and density contrasts and magnitude of the velocity dispersion
are quite similar in the 3D simulations. The clouds loss its identity
because they sweep up interstellar gas as they travel through the dense disk.
In order to visualize their effect on the disk, Figure~\ref{fig:planexy} shows
the density distribution across the equatorial plane ($z=0$) of the disk.
We see the presence of elongated regions of higher density that
resemble filaments and other more rounded regions of $\sim 300$ pc size,
which can be interpreted as holes.
The appearance of the density resembles that found in simulations
of supernova--driven turbulence.

The inclusion of cooling and diffuse UV heating will result in a larger
dynamical range in density. Nevertheless, we expect that the result
that the velocity dispersion of the warm phase is transonic is a
very general rule independent on these additional ingredients.
The most sensitive parameter in our simulations is the internal density
of the clouds. If this density is below a certain critical value, clouds
are destroyed in the upper halo and then the disk at $z\simeq 0$
cannot feel that the accretion is clumpy. In that case, the response
of the disk is similar to the ram pressure exerted by a smooth wind.
In order to circumvent this limitation, we plan to simulate the
accretion of clouds with an astrophysically motivated spectrum of 
masses and of sizes, so that they may have different internal densities.

\section{Discussion and conclusions}
The final fate of H\,{\sc i} gas filaments in the galactic halos 
is accretion onto the disk. 
It is important to have a view of the response of the gaseous
disk to accretion even if, surely, this is not the sole mechanism for
driving turbulent motions in the outer parts of galactic disks. 
Our simulations suggest that for mass accretion rates
consistent with current empirical determinations, 
the rain of CHVC onto the galactic
disks can maintain transonic turbulent motions in the warm phase.
In a forthcoming paper, we will report the evolution and amplification
of the magnetic field and vorticity, and compare the distributions of 
surface density and
velocity dispersion of our patch of the disk with observations.

\vskip 0.3cm
\noindent
This work has been partially supported from DGAPA--UNAM grant IN121609 and 
CONACyT project CB2006--60526.



\begin{thebibliography}{99}
\bibitem[Beckman et al.~(2003)]{bec03}
Beckman, J. E., L\'opez-Corredoira, M., Betancort-Rijo, J.,
Castro-Rodr\'{\i}guez, N., Cardwell, A. 2003, Ap\&SS, 284, 747
\bibitem[Booth \& Theuns~(2007)]{booth07}
Booth, C. M. \& Theuns, T. 2007, MNRAS, 381, 89
\bibitem[Dib \etal~2006]{Dib2006}
Dib, S., Bell, E., \& Burkert, A. 2006, ApJ, 638, 797
\bibitem[Kim et al.~1999]{kim99}
Kim, J., Ryu, D., Jones, T. W., \& Hong, S. S. 1999, ApJ, 514, 506
\bibitem[S\'anchez-Salcedo et al.~2007]{san07}
S\'anchez-Salcedo, F. J., Santill\'an, A., \& Franco J. 2007,
New Astronomy Reviews, 51, 104
\bibitem[Santill\'an et al.~1999]{santi99}
Santill\'an, A., Franco, J., Martos, M. \& Kim, J. 1999, ApJ, 515, 657
\bibitem[Santill\'an et al.~2007]{santi07}
Santill\'an, A., S\'anchez-Salcedo, F. J. \& Franco, J. 2007, ApJ, 662, L19
\bibitem[Stone \& Norman 1992a]{Stone1992a}
Stone, J.M., \& Norman, M. L. 1992a, ApJS, 80, 753
\bibitem[Stone \& Norman 1992b]{Stone1992b}
Stone, J.M., \& Norman, M. L. 1992b, ApJS, 80, 791
\end{thebibliography}
\end{document}